\documentclass[12pt]{iopart}

\usepackage{iopams}  
\usepackage{graphicx}
\usepackage{color}
\usepackage{varioref}

\begin{document}

\title[An analytic scaling relation for the maximum tokamak elongation against n=0 MHD resistive wall modes]{An analytic scaling relation for the maximum tokamak elongation against n=0 MHD resistive wall modes}
\author{Jungpyo Lee}
\address{MIT Plasma Science and Fusion Center, USA}
\author{Jeffrey Freidberg}
\address{MIT Plasma Science and Fusion Center, USA}
\author{Antoine Cerfon}
\address{NYU Courant Institute of Mathematical Science, USA}
\author{Martin Greenwald}
\address{MIT Plasma Science and Fusion Center, USA}
\ead{jungpyo@psfc.mit.edu}
\vspace{10pt}
\begin{indented}
\item[]November 2016
\end{indented}

\begin{abstract}
A highly elongated plasma is desirable in order to increase plasma pressure and energy confinement to maximize fusion power output. However, there is a limit to the maximum achievable elongation which is set by vertical instabilities driven by the $n=0$ MHD mode. This limit can be increased by optimizing several parameters characterizing the plasma and the wall. The purpose of our study is to explore how and to what extent this can be done. Specifically, we extend many earlier calculations of the $n=0$ mode and numerically determine scaling relations for the maximum elongation as a function of dimensionless parameters describing (1) the plasma profile ($\beta_p$ and $l_i$), (2) the plasma shape ($\epsilon$ and $\delta$), (3) the wall radius ($b/a$) and (4) most importantly the feedback system capability parameter $\gamma\tau_w$. These numerical calculations rely on a new formulation of $n=0$ MHD theory we recently developed [Freidberg et. al. 2015; Lee et. al. 2015] that reduces the 2-D stability problem into a 1-D problem. This method includes all the physics of the ideal MHD axisymmetric instability while reducing the computation time significantly, so that many parameters can be explored during the optimization process.     
The scaling relations we present include the effects of the optimal triangularity and the finite aspect ratio on the maximum elongation, and can be useful for determining optimized plasma shapes in current experiments and future tokamak designs.
\end{abstract}

%
%
%
%
%

\section{Introduction}
A highly elongated tokamak is desirable in order to increase plasma pressure and energy confinement, as verified in many experiments [1] and numerical simulations [2]. In the design of ITER, the expected confinement time $\tau_E$ was estimated by using experimentally derived empirical scaling relations. These relations, plus the well-known Troyon MHD beta limit, show a strong dependence on the elongation parameter $\kappa$ (i.e. $\tau_E\propto \kappa^{0.7}$ [3] and $\beta \propto (1+\kappa^2)$ [4]). The maximum value for the elongation is likely limited by axisymmetric ($n=0$) MHD resistive wall modes, which drive the vertical instability in the initial phase of the perturbation away from the equilibrium. As the vertical displacement becomes severe, finite toroidal modes ($n=1$, $n=2$, $\ldots$) may take over [5] and determine the evolution of the plasma during the disruption, which is often called a vertical displacement event (VDE), and has been studied in many tokamaks [6-10]. Since we are interested in configurations which avoid VDEs altogether (with the help of feedback stabilization), we focus on the $n=0$ mode in our work.  


There have been many numerical investigations of the $n=0$ MHD stability using different models (e.g. plasma surrounded by a perfectly conducting wall [11-13] or by a resistive wall [14-17]). However, these studies do typically not include the impact of the feedback system. In contrast, we have recently derived a variational formulation for the marginal linear stability of the $n=0$ mode which includes the presence of a thin resistive wall and which naturally integrates the effect of a realistic vertical instability control system through the introduction of a feedback parameter $\gamma\tau_w$ [17]. Along with this new formulation, we have developed a new numerical method [18] to efficiently compute the instability threshold. An important aspect of our numerical formulation is that it reduces the 2-D stability problem to an equivalent 1-D problem, which makes it computationally inexpensive, and allows us to explore multi-dimensional parameter space by running thousands of simulations, in order to obtain, through numerical curve fitting, useful analytic scaling relations. The main purpose of this article is to present these new scaling laws, which can be useful to optimize the plasma performance in existing tokamaks, and to design new machines.

A key additional feature of the present calculations as compared to our previous study [18], making our scaling relations more widely applicable, is the generalization of the plasma equilibrium pressure and current profiles. In [17, 18], the plasma profiles were restricted to the simple class of ``Solov'ev profiles" [19], which have the advantage of leading to MHD equilibria with explicit analytic representations, but have the caveat that they correspond to pressure and current profiles that are relatively flat radially as compared to typical experimental profiles. Specifically, the Solov'ev profiles have an internal inductance of about $l_i\simeq 0.4$, which is considerably smaller than the typical experimentally measured profiles characterized by $l_i > 0.7$. In order to compute equilibria for arbitrary plasma profiles, we use the Grad-Shafranov solver ECOM [20]. We also rely on ECOM to calculate, for each poloidal Fourier mode, the perturbed poloidal magnetic flux $\psi$ associated with the axisymmetric perturbation, as well as its normal derivative $\mathbf{n} \cdot \nabla \psi$. We remind the reader that following the notation introduced in [17], $\psi=\boldsymbol{\xi}_{\perp} \cdot \nabla \Psi$ where $\boldsymbol{\xi}_{\perp}$ is the perpendicular displacement vector, and $\Psi$ is the equilibrium poloidal flux, which satisfies the equilibrium Grad-Shafranov equation. The knowledge of the two quantities $\psi$ and $\mathbf{n} \cdot \nabla \psi$ at the plasma boundary is precisely what is required to apply our general formulation [18] and solve the equation $\delta W=0$. 
If $p$ is the number of poloidal Fourier modes used to decompose the perturbed flux $\psi$, the computational cost is about $p$ times larger than the corresponding cost for Solov'ev equilibria [18]. Even if so, our stability formulation still only requires solving two 1-D problems at the two radial interfaces (plasma-vacuum and  vacuum-wall), which is much more efficient than solving the full 2-D stability problem directly.

As described in [18], our methodology is the following. We first look for a set of parameters which satisfy the marginal stability condition $\delta W=0$ including the feedback control parameter $\gamma\tau_w$. Using this set of parameters, the maximum elongation $\kappa$ can be determined numerically in terms of the other parameters. In our studies, the maximum elongation is determined as a function of the following six critical dimensionless parameters: (1) beta poloidal ($\beta_p=4\int_V p d\mathbf{r}/ (\mu_0 I_\phi^2 R_0)$), (2) internal inductance ($l_i=2\int_V B_p^2 d\mathbf{r}/(\mu_0^2I_\phi^2R_0)$), (3) inverse aspect ratio ($\epsilon$), (4) triangularity ($\delta$), (5) the ratio of wall radius to the plasma radius ($b/a$) and (6) the feedback system performance parameter $\gamma\tau_w$. Here, the parameters ($\kappa$, $\delta$, and $\epsilon$) determining the shape of the plasma are as defined in [14]. For simplicity, we change the distance from the wall to the plasma simply by adjusting the parameter $\Delta_o$ defined by $b/a=(1+\Delta_o)$. In other words, we fix the shape of the wall relative to the shape of the plasma boundary by setting $\Delta_o=\Delta_i=(1/3)\Delta_v$, where $\Delta_o$, $\Delta_i$ and $\Delta_v$ are the outer, inner, and top gap between the plasma and the wall normalized by the minor radius $a$, respectively. This simple assumption is in agreement with the plasma and wall geometry of most existing tokamak experiments. Note that in our model, $\Delta_{o}$ determines the relation between the elongation and the triangularity of the plasma boundary and the elongation and the triangularity of the wall boundary: $\kappa_w=(\kappa+3\Delta_o)(1+\Delta_o)$ and $\delta_w=\delta(1+\Delta_o)$. 

The structure of the article is as follows. In Section 2, we describe how we use the equilibrium code ECOM to generalize our stability formulation [17,18] to physically relevant equilibrium pressure and current profiles. The fitting model of $\kappa(\beta_p,l_i,\epsilon,\delta,$ $b/a,\gamma\tau_w)$ based on our large number of simulations is presented in Sections 3 and 4, which highlight two important effects on the maximum elongation: the dependence on the optimal triangularity and the dependence on the aspect ratio. We summarize our results in Section 5, and highlight some remarkable features of the scaling laws.

\section{Implementation for arbitrary profiles}\label{Sec2}
In order to apply our formulation of the $n=0$ MHD stability problem [17,18] to arbitrary pressure and current profiles, we need to calculate the relation between $\psi$ and $\mathbf{n} \cdot \nabla \psi$ at the plasma boundary, which we call $\partial \Omega_p$, for these profiles. This relation is obtained by solving the neighboring equilibrium equation,
 \begin{eqnarray}
\Delta^{\star} \psi=-\left(\mu_0R^2\frac{d^2 p}{d \Psi^2}+\frac{1}{2}\frac{d^2 F^2}{d \Psi^2}\right)\psi \textrm{\;\;\;\;in\;} \Omega_p  \label{eq1}.
\end{eqnarray}
where $R$ is the radial coordinate in the $(R,\phi,Z)$ coordinate system associated with the tokamak geometry, $p$ is the plasma pressure, and $F(\Psi)=RB_{\phi}$ where $B_{\phi}$ is the toroidal magnetic field. In order to solve (\ref{eq1}), we first need to compute the Grad-Shafranov equation determining the equilibrium flux $\Psi$, with $p(\Psi)$ and $F(\Psi)$ given profiles, and the boundary condition $\Psi=0$ on $\partial\Omega_{p}$:
\begin{equation}
\Delta^{\star} \Psi =-\mu_{0}R^2\frac{dp}{d\Psi}-\frac{1}{2}\frac{d F^2}{d\Psi} 
\label{eqGS}
\end{equation}
We solve Eq. (\ref{eqGS}) with the axisymmetric equilibrium code ECOM [20]. Once $\Psi$ is known on the computational grid, so is the term in parenthesis in Eq. (\ref{eq1}). We can therefore also solve the linear Grad-Shafranov equation (\ref{eq1}) numerically using ECOM. The boundary conditions on $\psi$ are specified as follows.


The numerical formulation of the problem we presented in [18] relies on a Fourier series decomposition of the restriction of $\psi$ and $\mathbf{n} \cdot \nabla \psi$ to $\partial\Omega_p$ in terms of the poloidal arc-length variable $l$:
 \begin{eqnarray}
\psi=\sqrt{\frac{R}{R_0}}\sum_{n=1}^{p} a_n \sin(n l) \textrm{\;\;\;\;in\;} \partial\Omega_p, \\
\mathbf{n} \cdot \nabla \psi=\sqrt{\frac{R}{R_0}}\sum_{m=1}^{p} b_m \sin(m l) \textrm{\;\;\;\;in\;} \partial\Omega_p, 
\end{eqnarray}
where $l$ is the poloidal arc-length normalized to the range in $[0,2\pi]$ and $p$ is the number of poloidal modes. The relation between $\psi$ and $\mathbf{n} \cdot \nabla \psi$ then takes the form of the response matrix $\bold{T}$, whose component $T_{m,n}$ is defined by
 \begin{eqnarray}
b_m=\sum_{n=1}^{p} T_{m,n} a_n. 
\end{eqnarray} 
In the stability formulation presented in [18], which applies to Solov'ev profiles, we could derive an analytic expression for the response matrix, which took the form $(\bold{B}^{11})^{-1}(\bold{I}+\bold{A}^{11})$. To generalize the formulation for arbitrary profiles, one replaces that analytic matrix with numerically obtained response matrix $\bold{T}$.

Each row of the matrix $\bold{T}$ can be evaluated numerically by solving Eq. (\ref{eq1}) with the $\sin (nl)$ as the boundary condition on $\psi$ at $\partial \Omega_p$, and by using the solution to this equation to numerically evaluate $(\mathbf{n} \cdot \nabla \psi)_{n}(l)$ on $\partial \Omega_p$. Using the inverse Fourier series, one can find
 \begin{eqnarray}
T_{m,n} = \frac{1}{\pi}\int_0^{2\pi}{dl} {(\mathbf{n} \cdot \nabla \psi)}_{n}\sqrt{\frac{R_0}{R}}\sin (ml) \label{IFT}.
\end{eqnarray} 

In ECOM, the Grad-Shafranov equation is reexpressed as a nonlinear Poisson problem, which is solved iteratively. Solving the equilibrium Grad-Shafranov equation for $\Psi$ and Eq. (\ref{eq1}) for $\psi$ typically takes fewer than 10 iterations each. We call that number $n_{iter}$. Because we need to solve Eq. (\ref{eqGS}) once and Eq. (\ref{eq1}) $p$ times, the total cost to compute the response matrix $\bold{T}$ is $O((p+1)n_{iter}t_{\Omega})$, where $n_{iter}$ is the number of iterations for the convergence of the solver, and $t_{\Omega}$ is the time for each ECOM Poisson solve in $\Omega_p$. We calculate the coefficients of the response matrix via the Fast Fourier Transform, making the computational cost of this part of the stability calculation negligible.

In the next sections, we use the numerical formulation presented in this section to calculate the maximum elongation for several pressure and current profiles, corresponding to various $l_i$ and $\beta_p$. Specifically, we consider the class of pressure profiles, $\mu_0d{p}/d\bar{\Psi}=p_{0}(1-(1-\bar{\Psi})^{p_{in}})^{p_{out}}$, and $F$ profiles, $(1/2)(d{F}^2/d\bar{\Psi})=F_{0}(1-(1-\bar{\Psi})^{f_{in}})^{f_{out}}$, available in ECOM, where $\bar{\Psi}$ is the poloidal flux normalized to the interval $[0,1]$. In the present work, we keep The outer exponents $p_{out}=1.0$ and $f_{out}=1.0$ fixed for simplicity, and adjust the inner exponents $p_{in}$ and $f_{in}$ to have the intended profile shape for $l_i$, and the ratio $p_0/F_0$ to obtain the desired $\beta_p$.

Note that as compared to our previous study [18], we also have to recalculate the feedback parameter $\gamma \tau_w$ for the more realistic pressure and current profiles considered here. The results are given in Appendix A. The feedback parameter $\gamma \tau_w$ which was previously deduced [18] was obtained based on the assumption of Solov'ev profiles. These profiles correspond to a low value of inductance,  $l_i\sim0.4$, as compared to the typically observed values, $l_i\sim0.8$. The table given in Appendix A uses experimentally relevant $l_i$ and $\beta_p$. 


\section{The optimum triangularity}
It has been shown in [18] for low $l_i$ profiles that the maximum elongation can be increased by optimizing triangularity and the value of the optimal triangularity increases as the inverse aspect ratio $\epsilon$ increases. Figure 1 shows that the situation is more complicated for general profiles, and that the dependence of the elongation on the triangularity is a sensitive function of the value of $l_i$ and of $\beta_p$. In this section, we determine a scaling law for the optimal triangularity, and use this scaling law as well as numerical simulations to examine the conditions for the existence of an optimal triangularity.

\subsection{Model scaling laws}\label{sec2:1}
In the absence of a more educated guess, we model the dependence of $\kappa$ on $\delta$ by a quadratic form:
\begin{eqnarray}
\kappa=\kappa_{opt}-\kappa_{\delta} (\delta- \delta_{opt})^2 \label{kdquad}.
  \end{eqnarray}
The quadratic form chosen here can be viewed as the lowest order Taylor expansion of $\kappa$ around the optimal point $\delta_{opt}$, and is empirically justified by the curves $\kappa(\delta)$ we show in Figure 1. In (\ref{kdquad}) the optimal triangularity $\delta_{opt}$, the coefficient $\kappa_{\delta}$ as well as $\kappa_{opt}$ depend strongly on $\epsilon$, $l_i$, $\beta_p$, $\Delta_o$, and $\gamma\tau_w$. Intuitively, the existence of an optimal triangularity can be viewed as the result of competing effects between the pressure driven term and the line bending term in $\delta W$, which are sensitive to the plasma profiles. From our simulation results we have found that a good fit to the numerical data for the triangularity coefficients is given by
  \begin{eqnarray}
  \delta_{opt}&=& \hat{\delta} l_i^{\alpha_1} \beta_p^{\alpha_2} \epsilon^{\alpha_3}, \nonumber \\
  \kappa_\delta&=& \hat{\kappa}l_i^{\beta_1} \beta_p^{\beta_2} \epsilon^{\beta_3}, \label{deltaopt1}
\end{eqnarray}
where $\alpha_{1}$, $\alpha_{2}$, $\beta_{1}$, and $\beta_{2}$ are constants, $\alpha_3(l_i,\beta_p,\gamma \tau_w, \Delta_o)=\alpha_4+\alpha_5 l_i+\alpha_6\beta_p+\alpha_7(\gamma \tau_w)+\alpha_8(1+\Delta_o)$ and $\beta_3(l_i,\beta_p,\gamma \tau_w, \Delta_o)=\beta_4 +\beta_5 l_i+\beta_6\beta_p+\beta_7(\gamma \tau_w)+\beta_8(1+\Delta_o)$. Note the complicated dependence of the $\epsilon$ exponent on the plasma parameters. 

\begin{figure} 
\includegraphics[scale=0.4]{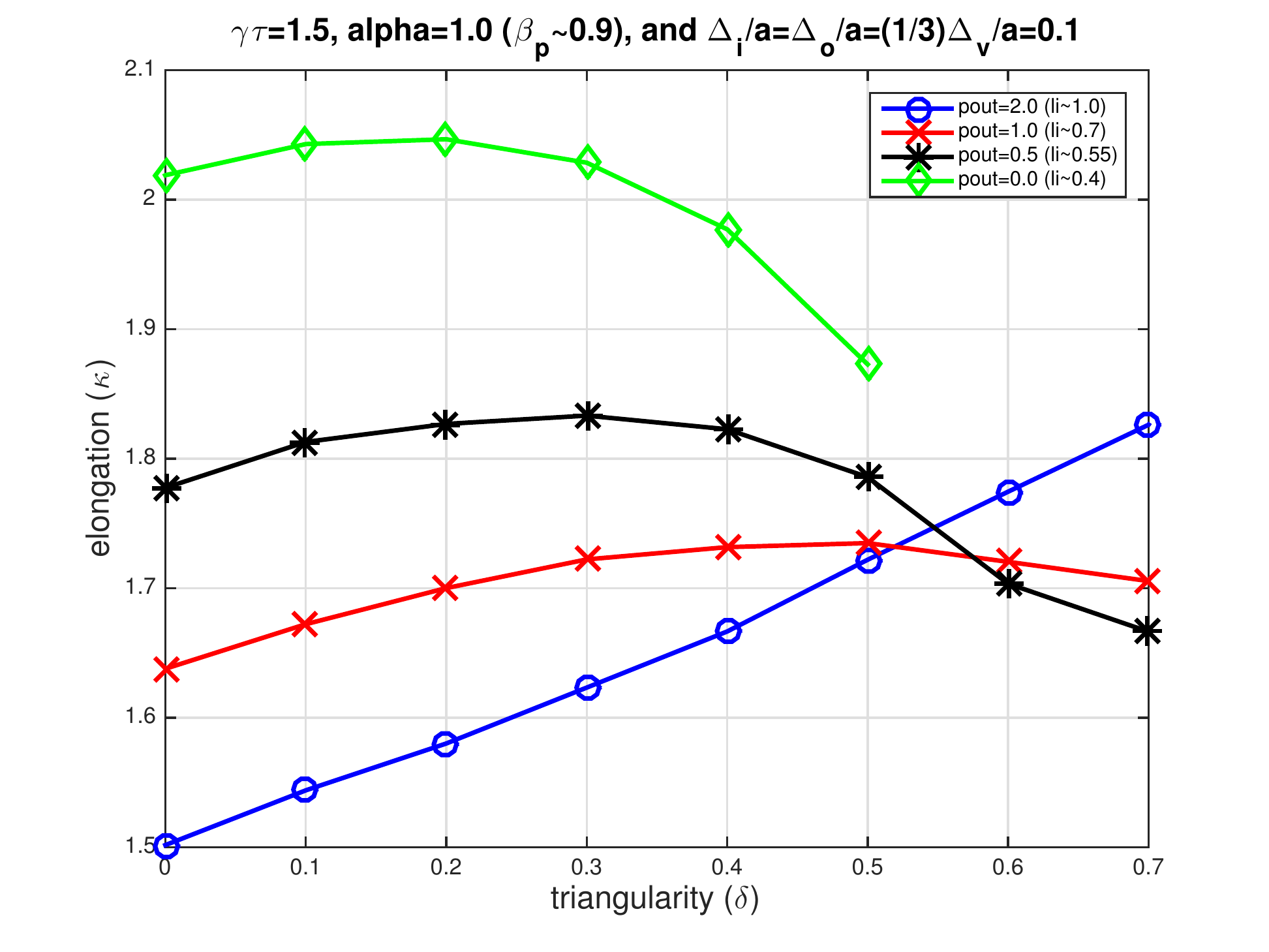}\includegraphics[scale=0.4]{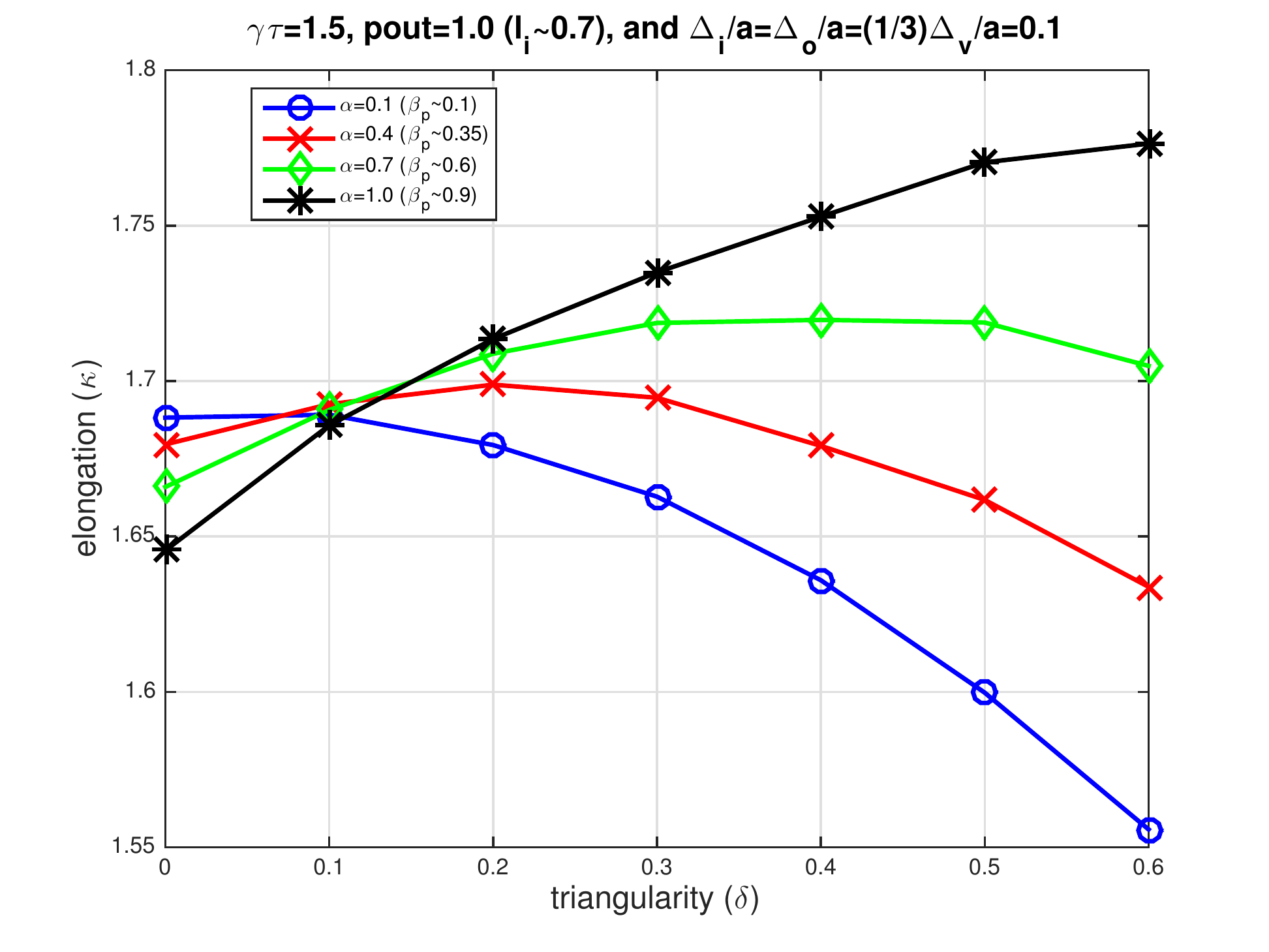}
\caption{Maximum elongation versus triangularity for various (a) internal inductance and (b) poloidal beta}
\label{fig:k_vs_d}
\end{figure}

\subsection{Fitting results}\label{sec2:2}
We calculate the numerical coefficients in the scaling law for the optimal triangularity by relying solely on our simulation data, in which the existence of an optimal triangularity is observed in the range $0\le\delta<0.8$. 
  Taking the log of the factors in Eq. (\ref{deltaopt1}) and using least squares fitting, we obtain the following best fit:
      \begin{eqnarray}
  \delta_{opt}= 2.30 l_i^{1.27} \beta_p^{-0.01} \epsilon^{(1.21 -0.76l_i-1.22\beta_p-0.001(\gamma \tau_w)+1.21(1+\Delta_o))},  \label{delta_res1}
  \end{eqnarray}
where the standard deviation of the fit is $\sigma=0.09$. As shown in Figure 2-(a), the difference between the scaling law and the numerical results is small for low to moderate $\delta$, and relatively high for high triangularity, i.e. $\delta_{opt}>0.5$. Mathematically, this can be explained by the fact that the curves corresponding to high optimal triangularity are flatter in the neighborhood of the optimum than the curves corresponding to low optimal triangularity, as can be seen in Figure \ref{fig:k_vs_d}. Physically, this is the signature of a complex interplay between the role of triangularity and the other relevant parameters in the $n=0$ linear MHD physics. Finally, some of the observed inaccuracy of the scaling law for large $\delta_{opt}$ is likely also due to the fact that ECOM is less accurate when the triangularity is relatively high.
  
The scaling law for the sensitivity coefficient $\kappa_\delta$ is rather complicated. For a wide range of parameters, we calculate $\kappa_\delta$ by computing $\kappa$ at $\delta=\delta_{opt}-0.2$, $\delta=\delta_{opt}$ and $\delta=\delta_{opt}+0.2$. A least squares fit of the resulting data set then yields
     \begin{eqnarray}
  \kappa_{\delta}&=&0.27 l_i^{-2.88} \beta_p^{0.10} \epsilon^{(0.45 +0.24l_i-0.23\beta_p+0.19(\gamma \tau_w)-0.75(1+\Delta_o))}  \label{C_res1}.
\end{eqnarray}
We see in Figure 2-(b) that the fit has a large standard deviation, which we calculate to be $\sigma=0.31$. This suggests that the simple quadratic model in Eq. (\ref{kdquad}) is in fact too simple to obtain an accurate fit. Even if so, $\kappa_{\delta}$ as given by Eq. (\ref{C_res1}) is a good indicator of the sensitivity of $\kappa$ on $\delta$. The formula in particular shows a robust tendency for $\kappa_{\delta}$ to decrease as $l_i$ increases. This can be explained as follows: as $l_i$ increases, the plasma current density is more concentrated in the core, and the effect of surface triangularity on the $n=0$ MHD mode is effectively reduced.



   \begin{figure} 
\includegraphics[scale=0.4]{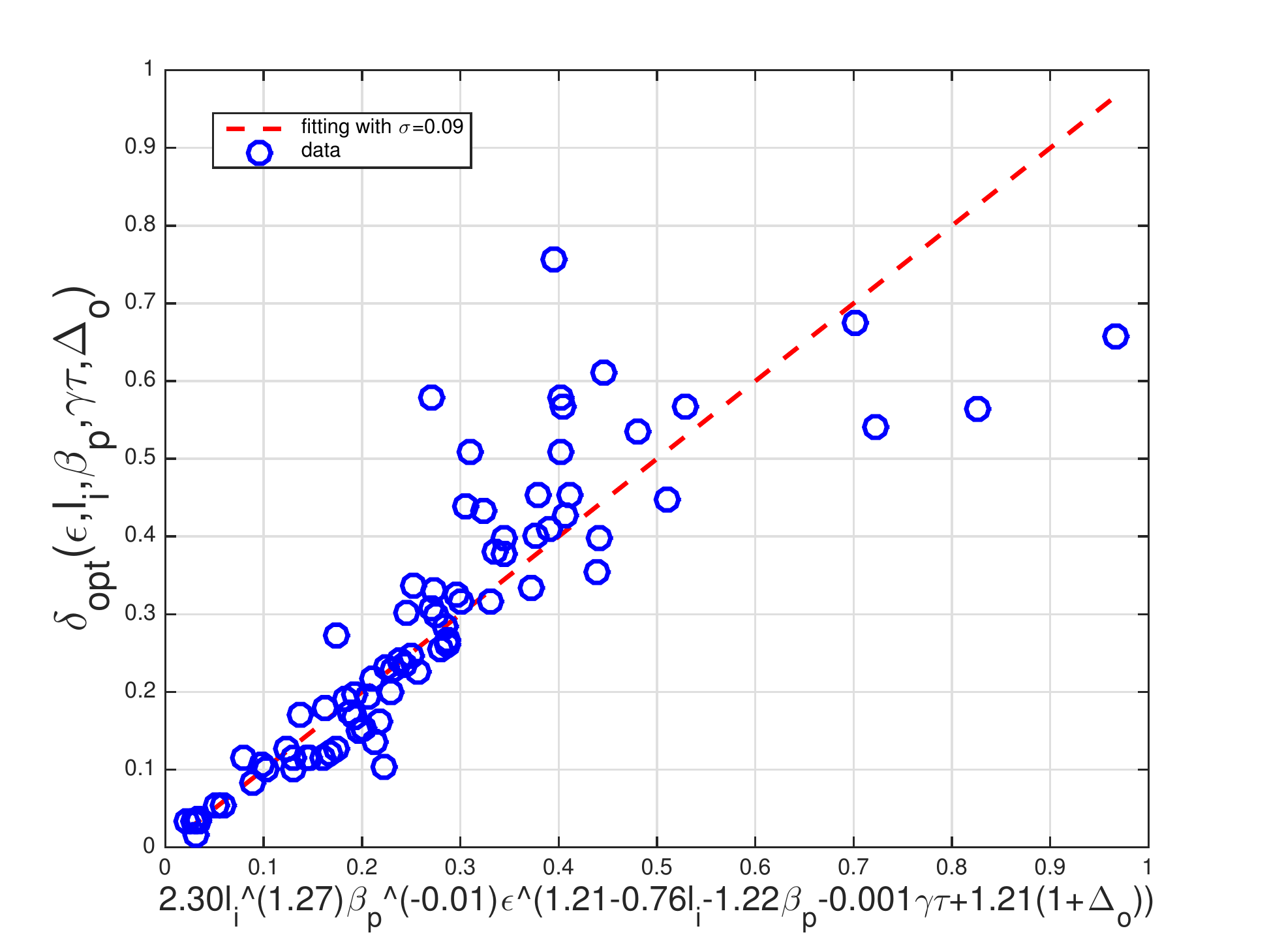}\includegraphics[scale=0.4]{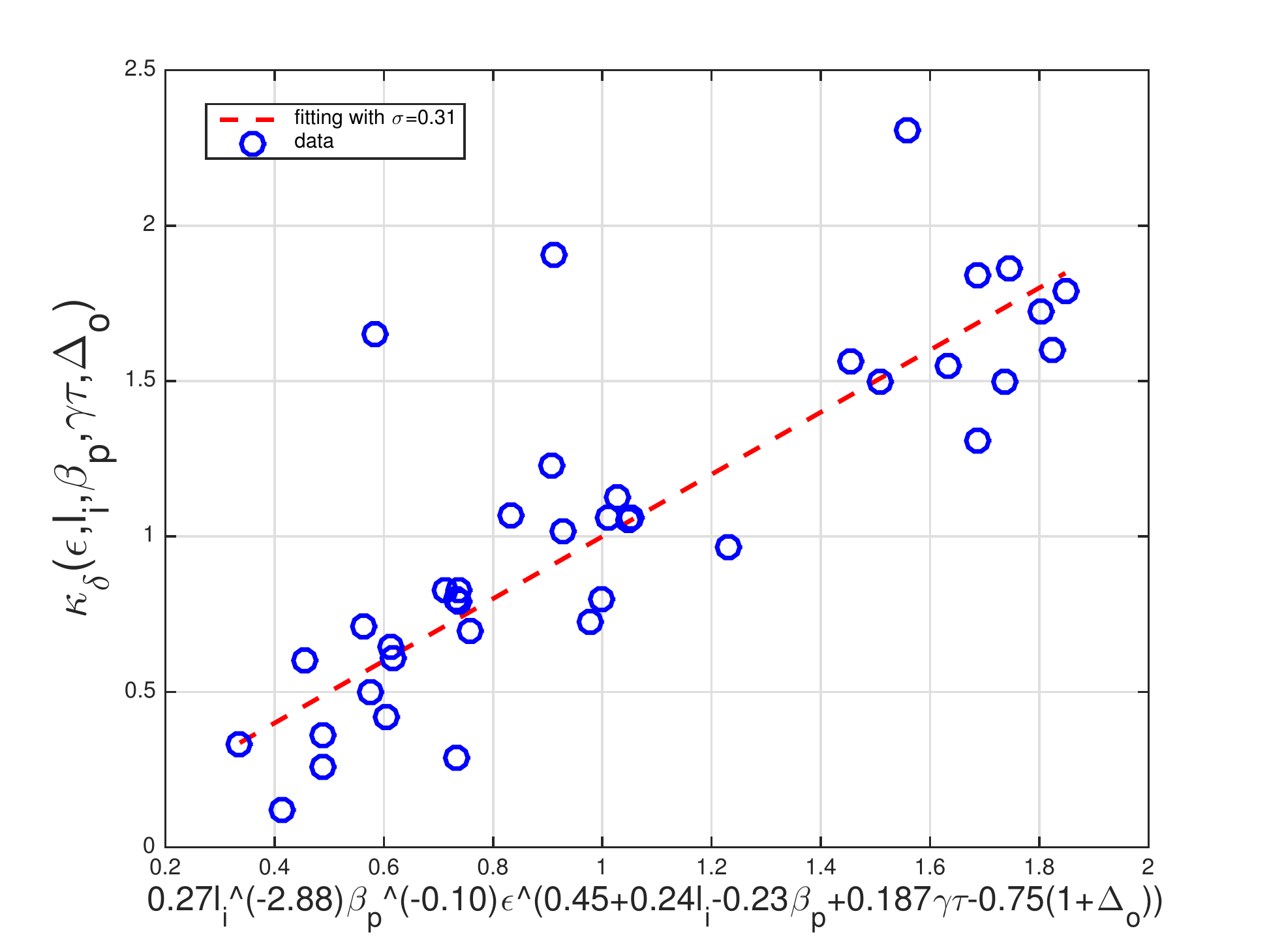}
\caption{Fitting of (a) $\delta_{opt}$  (b) $\kappa_\delta$ using the simulation results in which an optimal triangularity exists
}
\end{figure}

 \subsection{Condition for the existence of an optimal triangularity}\label{sec2:3}
 We have found that optimal triangularity generally increases as $\epsilon$, $l_i$ or $\beta_p$ increases. This is very likely due to the fact that the Shafranov shift increases when either of these three parameters increases. Note that the $\epsilon$ exponent in the scaling law in Eq. (\ref{delta_res1}) is approximately proportional to the theoretical value of Shafranov shift in the low $\epsilon$ limit: $-1.22 \beta_p-0.76l_i \sim  -1.22(\beta_p+0.5 l_i)$). This means that the optimal triangularity at the wall boundary needs to increase along with the Shafranov shift to stabilize the $n=0$ mode effectively. Because the Shafranov shift moves the core towards the low field side, the effective triangularity averaged over the total plasma volume increases due to the shift. The shift of the core increases with $\epsilon$, $l_i$ or $\beta_p$. 
 
Figure 3-(a) shows a contour plot of the optimal triangularity as a function of $\beta_p$ and $l_i$ obtained by running a large number of simulations, with a fixed $\epsilon=0.3$. Figure 3-(b) shows the equivalent figure as obtained from the scaling law for $\delta_{opt}$ in Eq. (\ref{delta_res1}). The two contour plots are reasonably well matched. Figure 3-(b) shows that sufficiently large values of $l_i$ or $\beta_p$ lead to a critically large value of $\delta_{opt}$. When this occurs (i.e. $\delta_{opt}>\sin({1.0}) =0.84$), the well-known Miller cross section [21] used in our simulations breaks down in the sense that the plasma shape is no longer convex and assumes a bean shape instead, which is not relevant for current tokamaks, and unlikely for future tokamaks [22]. In practice, it is therefore reasonable to limit triangularity to $\delta\leq 0.7$. Hence, we say that if the optimal triangularity $\delta_{opt}$ according to Eq. (\ref{delta_res1}) is comparable to or larger than the value 0.7, there is no optimal triangularity. In that case, the maximum achievable elongation is obtained by maximizing the triangularity. The white region in the upper half corner of Figure 3-(a) corresponds to a region in $\beta_{p}$-$l_{i}$ space in which no optimal triangularity was found numerically in the range $0\leq\delta_{opt}\leq 0.7$. The white region in the lower half corner of Figure 3-(a) for the small $\beta_p$ and $l_i$ is not simulated because this parameters are not relevant for current tokamaks. As a rule of thumb, the optimal triangularity tends to exceed 0.7 if $l_i\geq0.9$ or $\beta_p\geq1.4$ for $\epsilon=0.3$, $\gamma \tau_w=1.5$ and $\Delta_o=0.1$.  
 
   \begin{figure}[h]
   \begin{center}
(a)\;\;\;\;\;\;\;\;\;\;\;\;\;\;\;\;\;\;\;\;\;\;\;\;\;\;\;\;\;\;\;\;\;\;\;\;\;\;\;\;\;\;\;\;\;\;\;\;\;\;\;\;\;\;\;\;\;\;\;\;\;\;\;\;(b)\;\;\;\;\;\;\;\;\;\;\;\;\;\;\;\;\;\;\;\;\;\;\;\;\;\;\;\;\;\;\;\;\;\;\;\;\;\;\;\;\;\;\;\;\;\;\;\;\;\;\;\;\;\;\;\;\;\;\;\;\\
\includegraphics[scale=0.42]{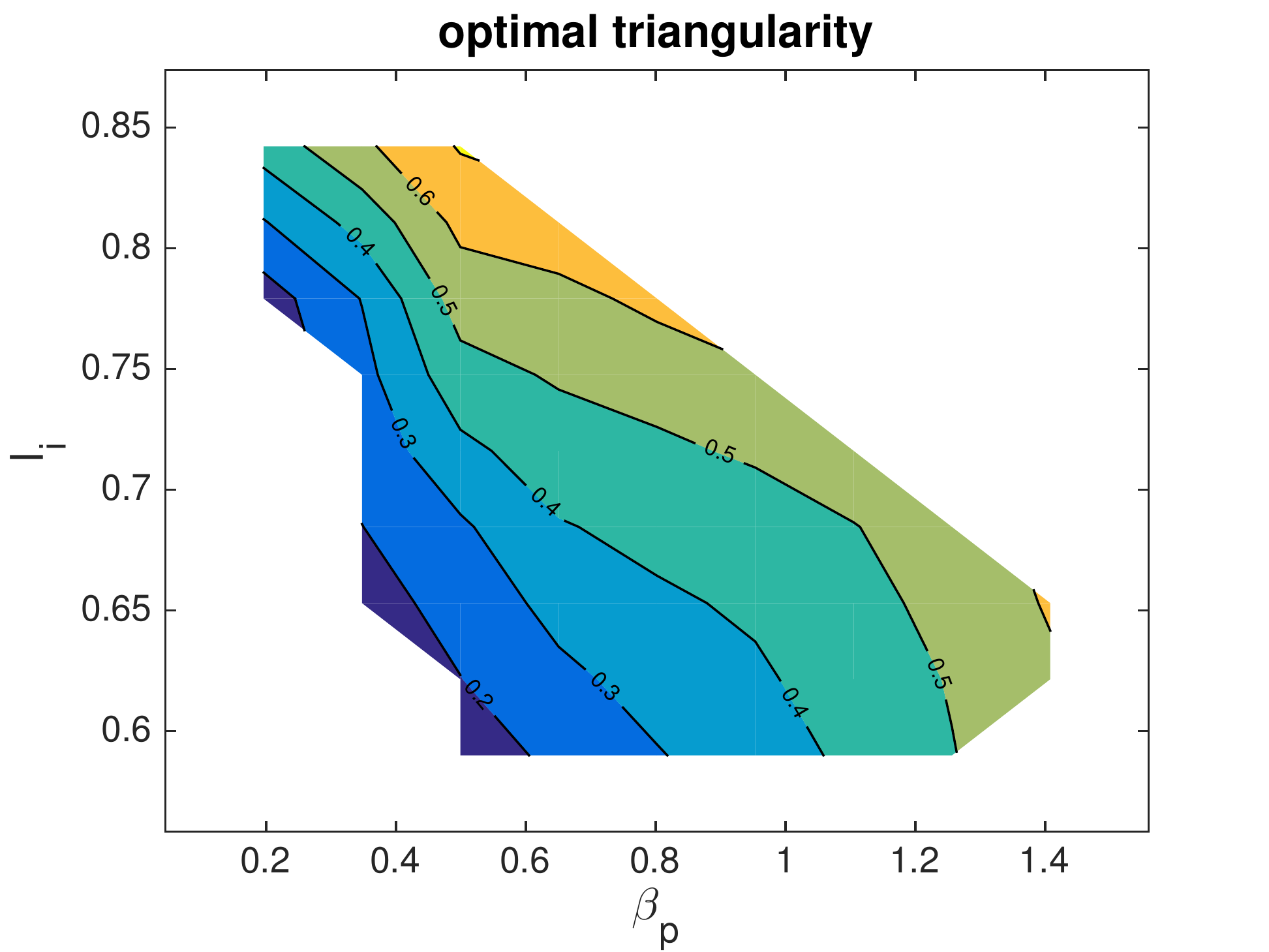}\includegraphics[scale=0.4]{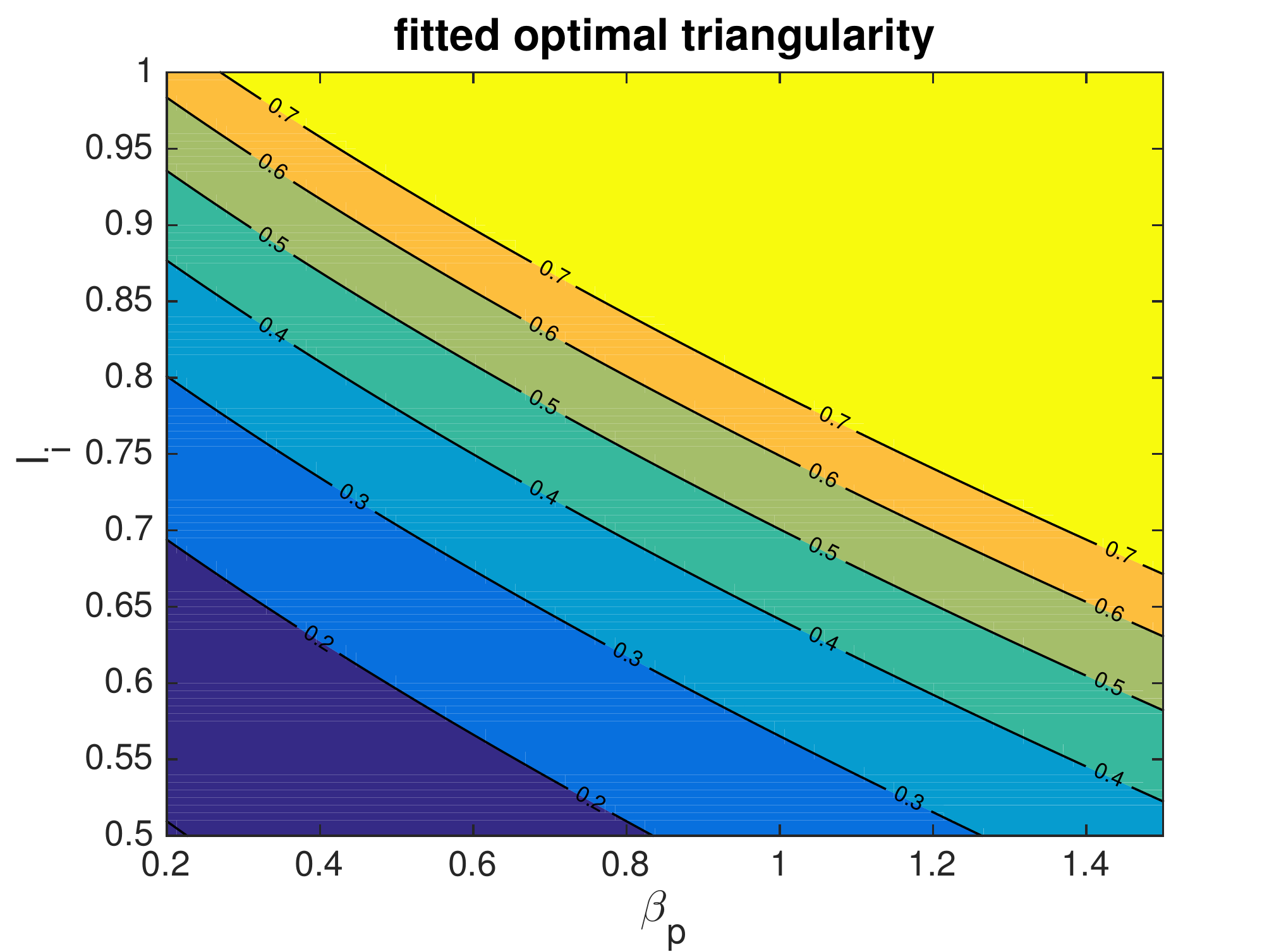} 
\caption{Contour of optimal triangularity $\delta_{opt}$ in terms of $l_i$ and $\beta_p$ using a) simulation result and (b) fitted formula in Eq. (\ref{delta_res1}) for $\gamma \tau_w=1.5$, $\Delta_o=0.1$ and $\epsilon=0.3$. The upper-right white space in (a) corresponds to a region in $\beta_p$ - $l_i$ parameter space having no optimal triangularity in the simulations.
}   
   \end{center}
   \end{figure} 
 

\section{Dependence on the aspect ratio}\label{sec1:tri}

The dependence of the maximum elongation on the aspect ratio $\epsilon$ has been investigated previously [11]. We found in our simulations that this dependence is far more complicated than what the results from previous work would indicate. This is because of the dependence of the effective triangularity on the aspect ratio, as discussed in Section 3. At very large aspect ratio, the optimal triangularity is very low, in which case we recover previous results, as we will show. However, the situation is more subtle at finite $\epsilon$, when $\delta$ effects come in. This is the central point of this section.
 
Let us start with the situation corresponding to very large aspect ratio. Figure 4 shows the simulation results of the maximum elongation for a large aspect ratio ($\epsilon=0.01$) device in terms of the internal inductance for various values of the triangularity. For such a large aspect ratio, $\delta_{opt}\simeq0.0$, and for this optimal triangularity (corresponding to the black curve), the maximum elongation increases as $l_i$ decreases, as found in previous studies [11], because lower $l_i$ means a smaller effective plasma-wall distance. However, as the triangularity increases away from the optimal triangularity, the maximum elongation is reduced, and the reduction is particularly significant at low $l_i$ when the effective distance between the plasma and the wall is small. We will return to the large aspect ratio case in Section 4.2, where we give an explicit dependence for the elongation on the physical parameters at very low $\epsilon$.
 
We determine below separate scaling laws for four different values of triangularity ($\delta=0.0$, $\delta=0.33$. $\delta=0.5$, and $\delta=0.7$). We do not include the dependence on triangularity in the scaling laws for two reasons. First, as we have seen with the relatively high variance associated with the scaling law for $\kappa_{\delta}$ in Eq. (\ref{kdquad}) and Eq. (\ref{C_res1}), it is challenging to construct scaling laws in terms of $\delta$ which are robust over a wide range of $\delta$ values. Moreover, as we have shown in Figure 4, the maximum elongation changes significantly as the triangularity deviates from $\delta_{opt}$, and the optimal triangularity depends sensitively on several parameters as shown in Eq (\ref{delta_res1}). Second, the scaling law for the maximum elongation at the optimal triangularity may not be directly useful to the fusion community because the triangularity in many experiments or tokamak designs is not only determined by the optimal triangularity with respect to the $n=0$ instability but also by other constraints and performance goals, such as the locations of the coils, turbulent transport, and other MHD instabilities [18]. As a result, most machines have a moderate triangularity (e.g. $\delta=0.33$ for ITER), while the optimal triangularity regarding the $n=0$ mode is likely to be too large ($\delta>0.7$) for typical experimental values of the parameters $l_i$, $\beta_p$, and $\epsilon$.
 
  \subsection{Model scaling laws}\label{sec3:1}

Our numerical results show that the maximum elongation can be accurately modeled by a simple form,
 \begin{eqnarray}
  \kappa&=&\kappa_0+\kappa_1\left(\frac{2\epsilon}{1+\epsilon^2}\right)^2,\label{kappam1}
\end{eqnarray}
where the quantities $\kappa_{0}$ and $\kappa_{1}$ depend on $\gamma\tau_{w}$, $l_{i}$, $\Delta_{0}$, and $\beta_{p}$ through the fitting formulae in Eqs. (\ref{kappa0}) and (\ref{kappa1}), but do not depend on $\epsilon$. In the next two paragraphs, we give physical explanations for the good fit between the simple form given by Eq. (\ref{kappam1}) and our numerical results, many of which confirm results obtained in previous studies [15,23].

Consider first the coefficient $\kappa_0$ which represents the maximum elongation in the limit of large aspect ratio. This coefficient is due to the effects of the finite distance between the plasma and the wall, which are independent of the magnitude of the aspect ratio [15]. In the limit in which the wall is at infinity, the optimal shape approaches a circle corresponding to $\delta_{opt}\rightarrow 0$ and $\kappa_0=1$. Mathematically, the wall can be moved to infinity in several ways: $\Delta_o\rightarrow \infty$, $l_i  \rightarrow \infty$, and $\gamma \tau_w\rightarrow 0$. For finite values of these parameters and a fixed triangularity, a good fit to the numerical simulations is obtained by assuming that $\kappa_0$ scales as
 \begin{eqnarray}
  \kappa_0&=&1.0+\hat{\kappa}_0  \frac{(\gamma \tau_w)^{\nu_1}}{ l_i^{\nu_2}  (1+\Delta_o)^{\nu_3}}.  \label{kappa0}
\end{eqnarray}
where $\hat{\kappa}_0$, $\nu_{1}$, $\nu_{2}$, and $\nu_{3}$ are scalar constants which will be computed through a fitting procedure and given in section 4.2. Note that the dependence of $\kappa_0$ on $\beta_p$ is very weak and can be ignored with a minimal loss in accuracy.

The aspect ratio dependence of the maximum elongation is determined by the coefficient $\kappa_1$ and the functional dependence on $\epsilon$ assumed in Eq. (\ref{kappam1}). Observe that as $\epsilon \rightarrow 0$, the maximum elongation is proportional to $\epsilon^2$ as expected from calculations of the natural elongation of a tokamak in Eq. (33) of [15] and Eq. (87) of [23]. Also, as $\epsilon \rightarrow 1$, the maximum elongation saturates. The rate of saturation depends on the parameters $l_i$ and $\beta_p$, as shown in Figure 5. A large Shafranov shift for a high $l_i$ or $\beta_p$ results in the increase of the effective triangularity of the plasma and a reduction of the maximum elongation because of the large difference between the optimal triangularity and the given triangularity. That being said, we will ignore for simplicity the dependence of the rate of saturation on $l_i$ or $\beta_p$, and fix the saturation rate in our model scaling law Eq. (\ref{kappam1}) by setting $d\kappa/d\epsilon=0$ at $\epsilon=1$. As we will show, this simplifying assumption leads to a reasonably accurate model.
 
Our numerical simulations show that an accurate scaling law for $\kappa_1$ can be written as 
  \begin{eqnarray}
  \kappa_1&=&\hat{\kappa}_1 l_i^{\mu_1} \beta_p^{\mu_2}(\gamma \tau_w)^{\mu_3}  (1+\Delta_o)^{\mu_4}.  \label{kappa1}
\end{eqnarray}
where $\hat{\kappa}_1$, $\mu_{1}$, $\mu_{2}$, $\mu_{3}$, and $\mu_{4}$ are scalar constants which will be computed through a fitting procedure and given in section 4.3.
    \begin{figure} 
\includegraphics[scale=0.30]{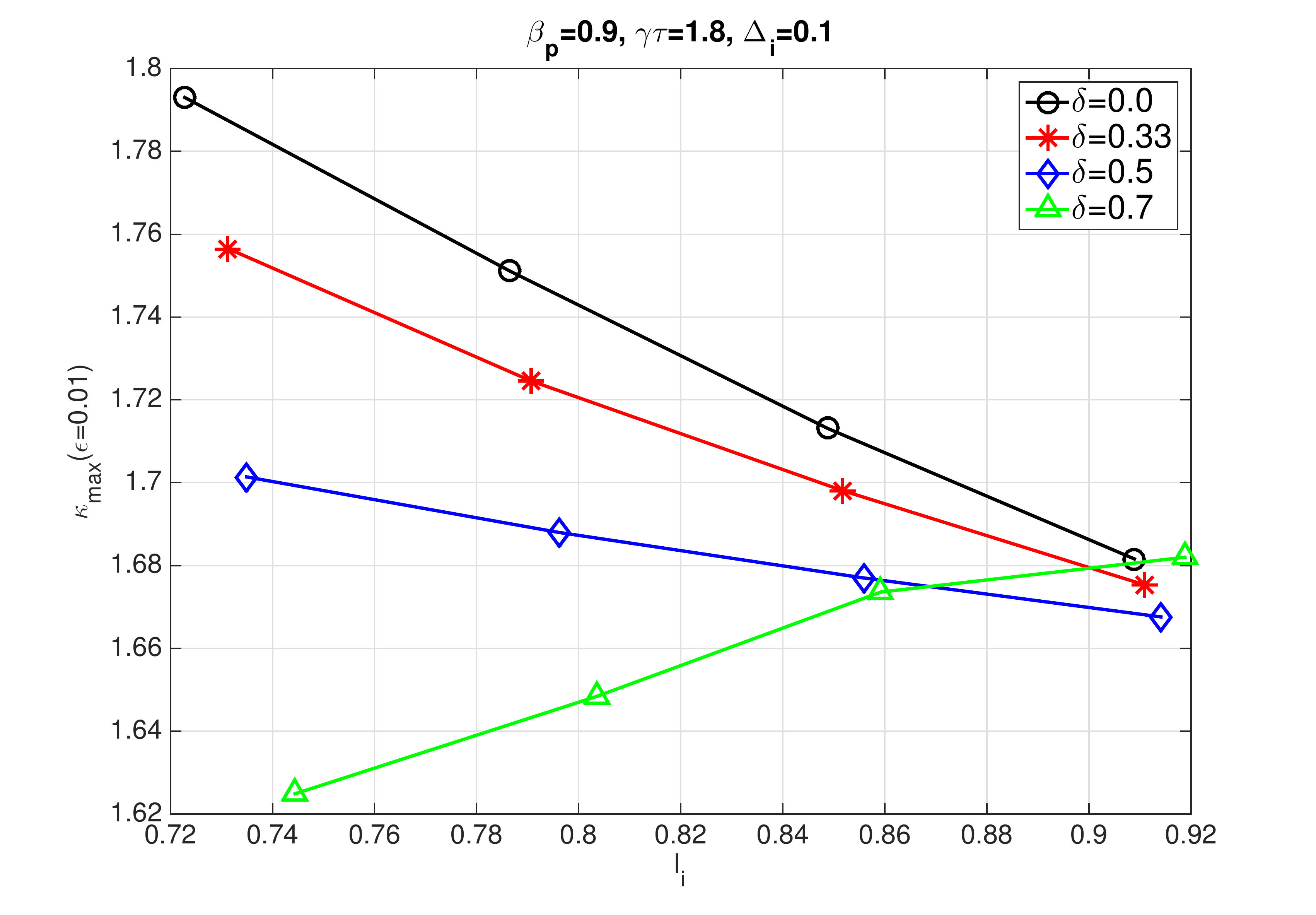}
\caption{$\kappa$ vs. $l_i$ with various $\delta$ for $\epsilon=0.01$, $\gamma \tau_w=1.8$ and $\Delta_o=0.1$. 
}
\end{figure}
    \begin{figure} 
\includegraphics[scale=0.35]{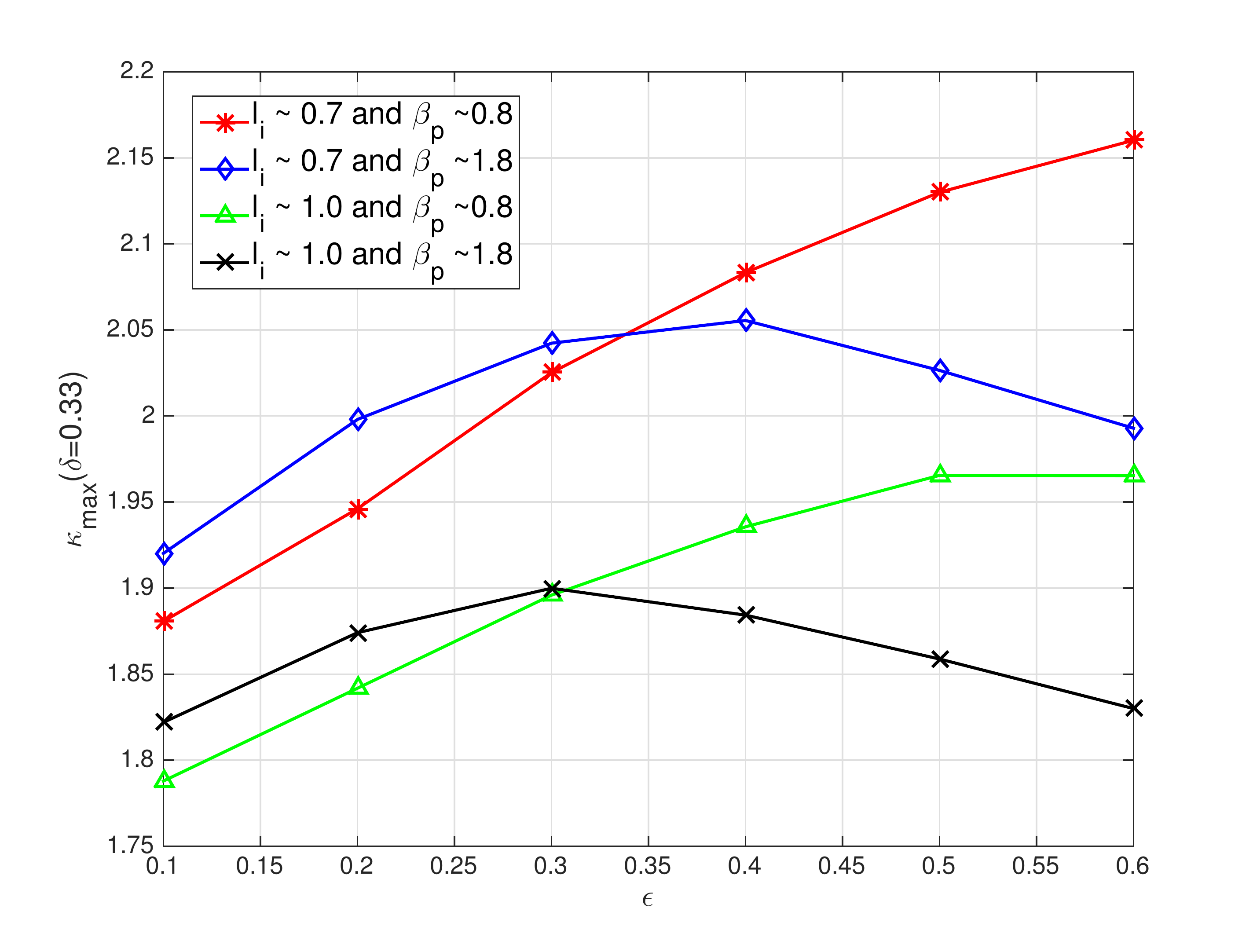}
\caption{$\kappa$ vs. $\epsilon$ for various $l_i$ and $\beta_p$, with $\delta=0.33$, $\gamma \tau_w=1.5$ and $\Delta_o=0.1$ fixed. }
\end{figure}


 \subsection{Fitting of $\kappa_0$}
As shown in Figure 6 for the case $\delta = 0.0$, the coefficients in the scaling relation for $\kappa_0$ in Eq. (\ref{kappa0}) are calculated by fitting the simulation results of $\kappa$ at $\epsilon=0.01$. For the four different values of the triangularity, the results are  
   \begin{eqnarray}
    \kappa_0&=&1.0+0.54 l_i^{-0.68} (\gamma \tau_w)^{0.62} (1+\Delta_o)^{-3.52}  \textrm{\;\;for\;$\delta=0.0$},\nonumber\\
        \kappa_0&=&1.0+0.54 l_i^{-0.47} (\gamma \tau_w)^{0.71} (1+\Delta_o)^{-4.00}  \textrm{\;\;for\;$\delta=0.33$}, \nonumber\\
            \kappa_0&=&1.0+0.55 l_i^{-0.08} (\gamma \tau_w)^{0.82} (1+\Delta_o)^{-4.74}  \textrm{\;\;for\;$\delta=0.50$}, \nonumber\\
                \kappa_0&=&1.0+0.63 l_i^{1.20} (\gamma \tau_w)^{1.14} (1+\Delta_o)^{-6.67}  \textrm{\;\;for\;$\delta=0.70$}, \label{kappa0_res0} 
\end{eqnarray}
where the standard deviation of the fitting is quite low ($\sigma=0.003, 0.006, 0.01,$ and $0.04$ for $\delta=0.0, 0.33, 0.5,$ and $0.7$, respectively). The observed $l_i$ dependence in Figure 4 is reflected in the increasing exponent of $l_i$ for larger values of the triangularity. Additionally, the absolute values of the exponents of $\gamma \tau_w$ and $(1+\Delta_o)$ increase as the triangularity increases. We observe that the dependence of the maximum elongation on the wall and feedback system becomes strong for larger values of the triangularity.

       \begin{figure} 
\includegraphics[scale=0.5]{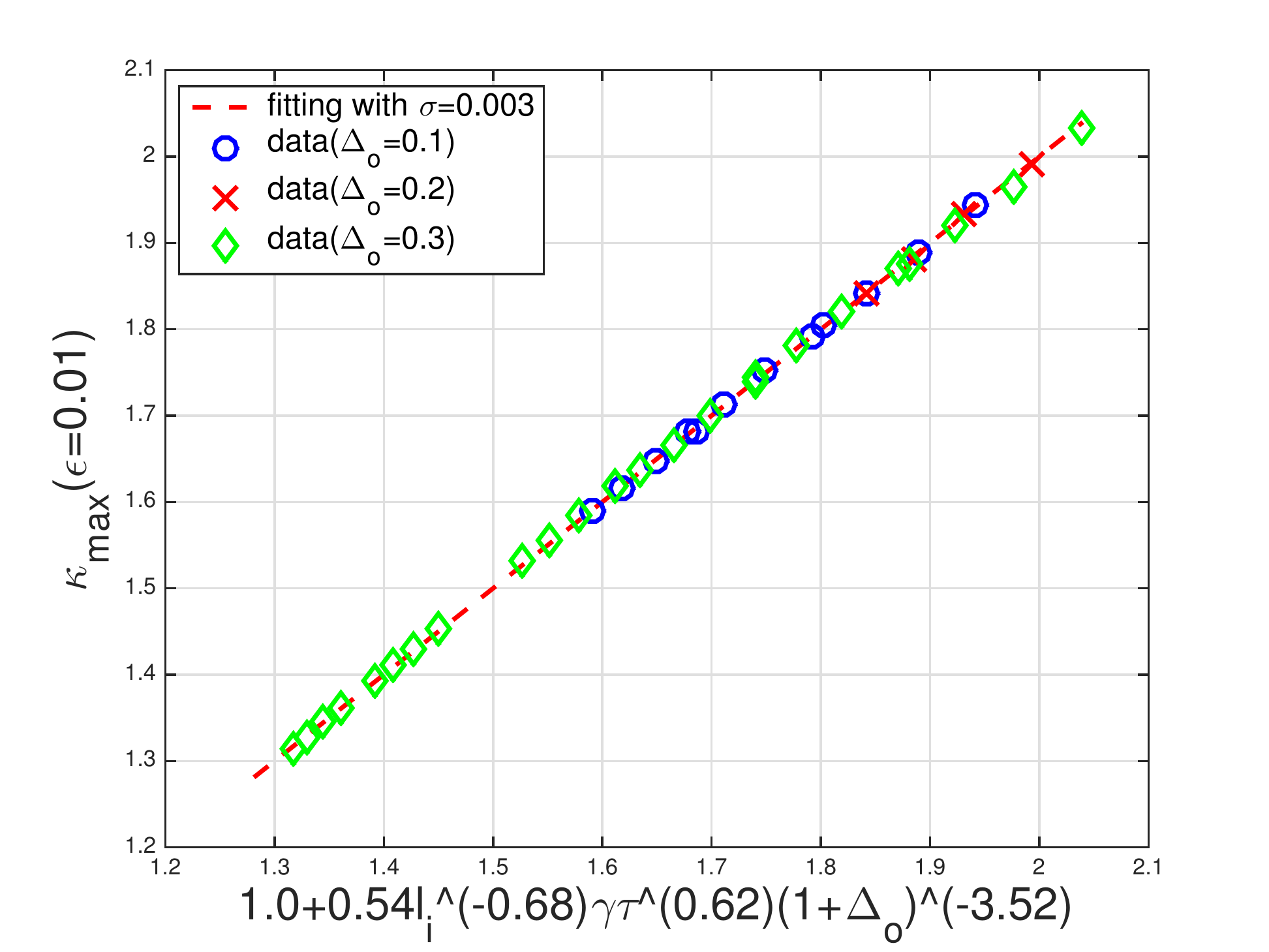}
\caption{Fitting of $\kappa_0$ using $\kappa$ at $\epsilon=0.01$ and $\delta=0.0$
}
\end{figure}
  \subsection{Fitting of $\kappa_1$} 
   The coefficient for the $\epsilon$ dependence, $\kappa_1$ in Eq. (\ref{kappa1}), can be estimated from the difference between $\kappa(\epsilon=0.01)$ and $\kappa(\epsilon=0.6)$, where  $\epsilon=0.6$ is the maximum value we considered here, as we experienced difficulties with our numerical code for higher values of $\epsilon$ and high triangularity.  For various values of the triangularity, $\kappa_{1}$ is given by
    \begin{eqnarray}
    \kappa_1&=&0.04 l_i^{-6.98} \beta_p^{-2.67}(\gamma \tau_w)^{-1.47} (1+\Delta_o)^{1.84}  \textrm{\;\;for\;$\delta=0.00$} ,\nonumber\\
      \kappa_1&=&0.35 l_i^{-1.42} \beta_p^{-0.04}(\gamma \tau_w)^{-0.27} (1+\Delta_o)^{0.42} \textrm{\;\;for\;$\delta=0.33$} , \nonumber\\
            \kappa_1&=&0.41 l_i^{-1.21} \beta_p^{0.06}(\gamma \tau_w)^{-0.18} (1+\Delta_o)^{0.68}   \textrm{\;\;for\;$\delta=0.50$},\nonumber\\
                \kappa_1&=&0.52 l_i^{-2.00} \beta_p^{0.17}(\gamma \tau_w)^{-0.50} (1+\Delta_o)^{2.32}  \textrm{\;\;for\;$\delta=0.70$} \label{kappa0_res1},
\end{eqnarray}
where the standard deviation of the fitting is reasonably low ($\sigma=0.01, 0.01, 0.03,$ and $0.06$ for $\delta=0.0, 0.33, 0.5,$ and $0.7$, respectively). 

Using scaling laws for $\kappa_0$ and $\kappa_1$ in Eq. (\ref{kappa0_res0}) and Eq. (\ref{kappa0_res1}), the simple scaling law for $\kappa$ in Eq. (\ref{kappam1}) leads to a good fit for all simulation results for parameters $\epsilon$, $l_i$, $\beta_p$, $\Delta_o$, and $\gamma\tau_w$ varied over a wide range. Figures 7-10 illustrate this remarkably good agreement, with standard deviations $\sigma=0.02, 0.05, 0.08,$ and $0.14$ for $\delta=0.0, 0.33, 0.5,$ and $0.7$, respectively. 
   
       \begin{figure} 
   \begin{center}
\includegraphics[scale=0.4]{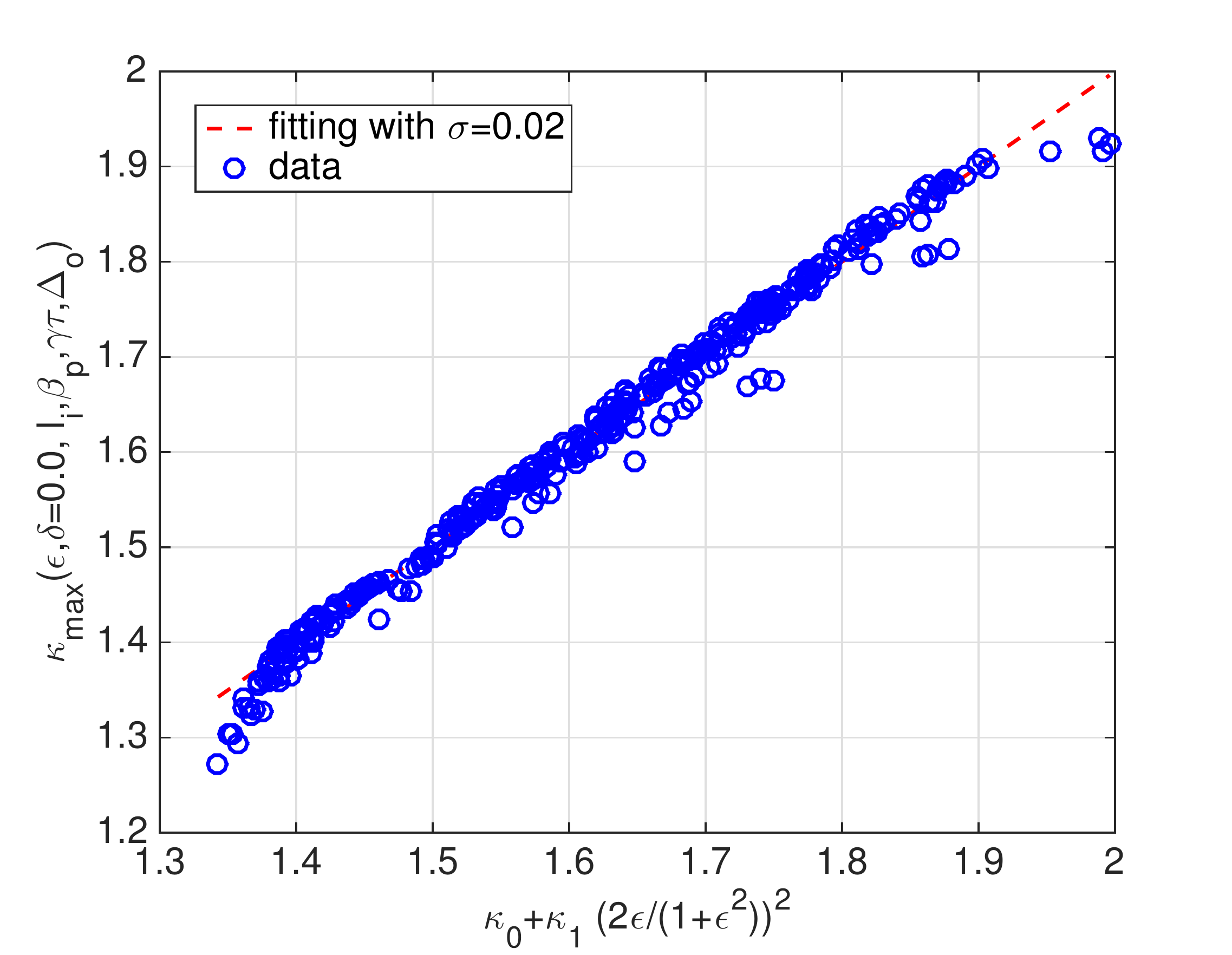}
\caption{Fitting of $\kappa$ for $\delta=0.0$ using Eq. (\ref{kappam1}) with $\kappa_0$ and $\kappa_1$ given by Eq. (\ref{kappa0_res0}) and Eq. (\ref{kappa0_res1})}
   \end{center}
   \end{figure} 
          \begin{figure} 
   \begin{center}
\includegraphics[scale=0.4]{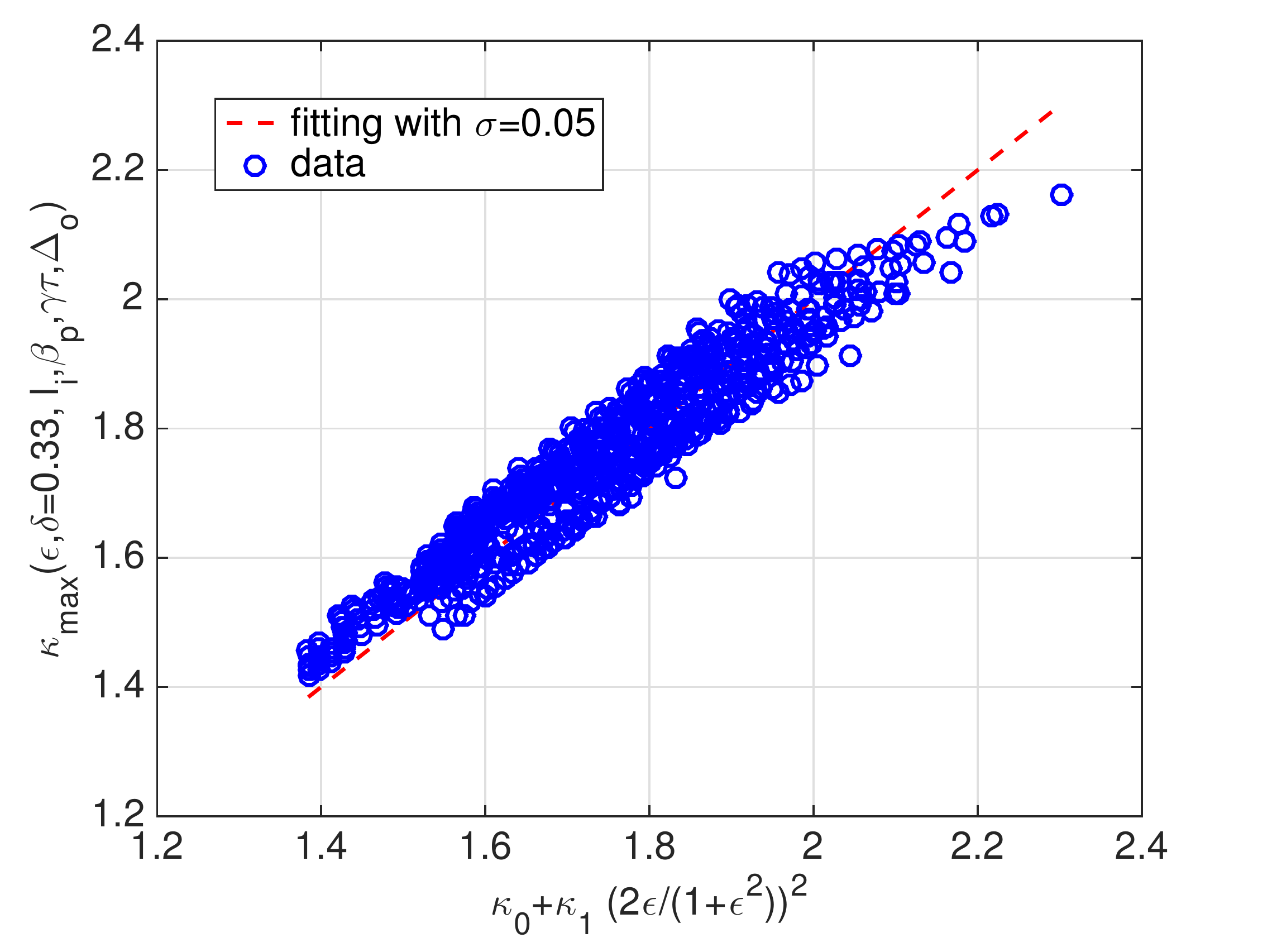}
\caption{Fitting of $\kappa$ for $\delta=0.33$ using Eq. (\ref{kappam1}) with $\kappa_0$ and $\kappa_1$ given by Eq. (\ref{kappa0_res0}) and Eq. (\ref{kappa0_res1})}
   \end{center}
   \end{figure} 
          \begin{figure} 
   \begin{center}
\includegraphics[scale=0.4]{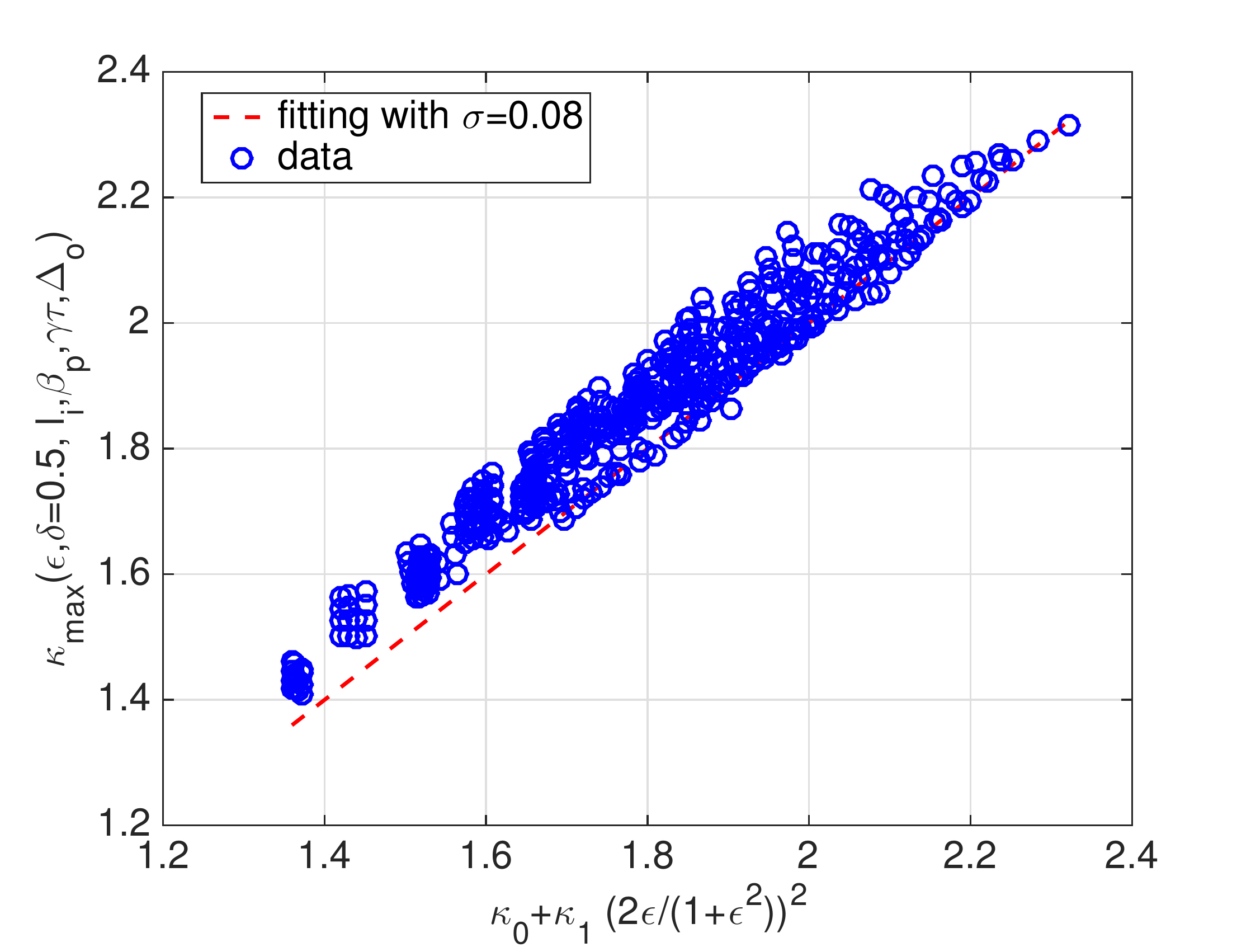}
\caption{Fitting of $\kappa$ for $\delta=0.5$ using Eq. (\ref{kappam1}) with $\kappa_0$ and $\kappa_1$ given by Eq. (\ref{kappa0_res0}) and Eq. (\ref{kappa0_res1})}
   \end{center}
   \end{figure} 

          \begin{figure} 
   \begin{center}
\includegraphics[scale=0.4]{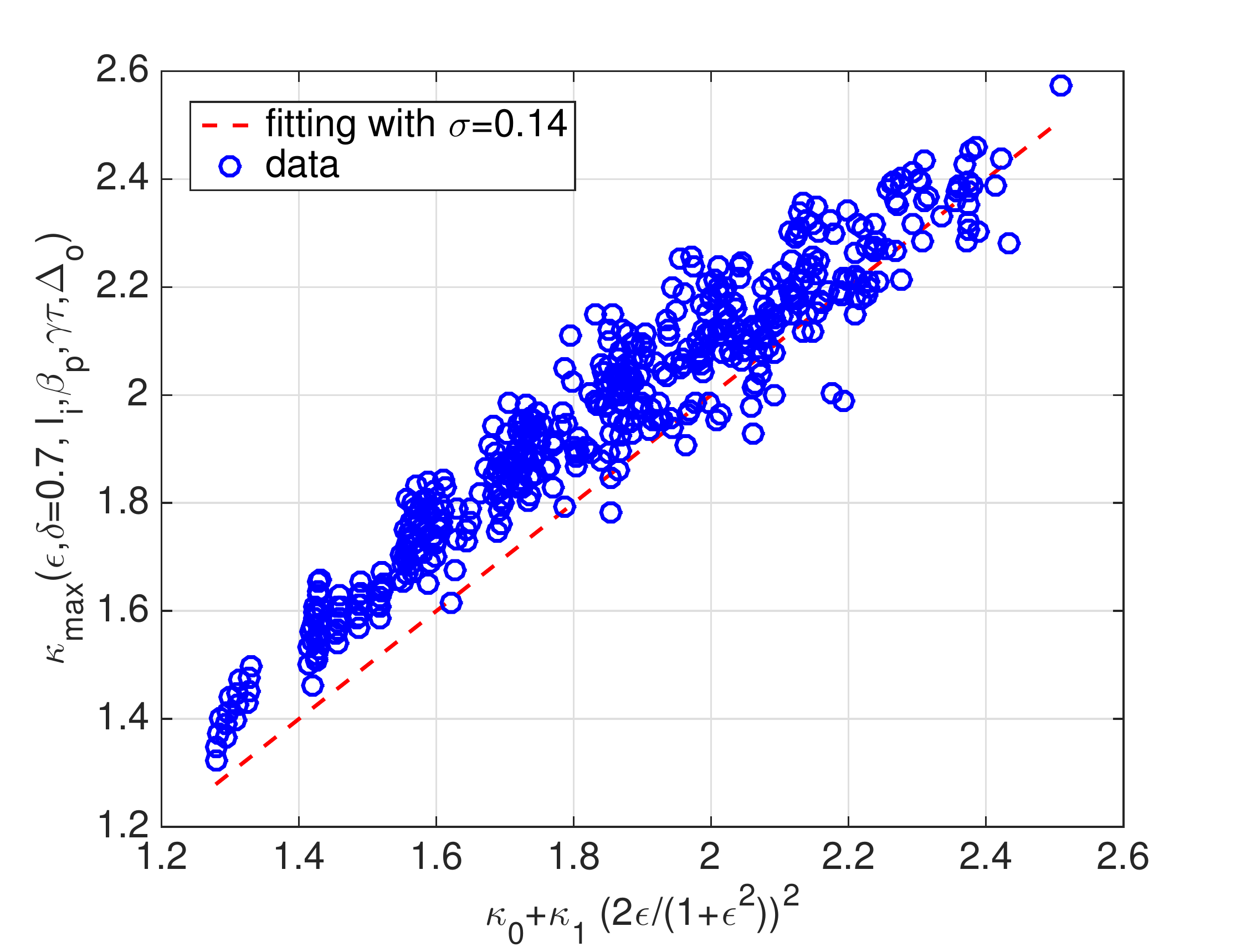}
\caption{Fitting of $\kappa_{max}$ for $\delta=0.7$ using Eq. (\ref{kappam1}) with $\kappa_0$ and $\kappa_1$ given by Eq. (\ref{kappa0_res0}) and Eq. (\ref{kappa0_res1})}
   \end{center}
   \end{figure}




 \section{Discussion}
We applied a new and fast numerical scheme to a recently developed variational formulation [17,18] to compute the maximum achievable elongation in tokamaks in the presence of a resistive wall and a feedback stabilization system. The speed of our numerical solver allowed us to explore a wide range of parameter space, and derive analytic scaling laws for the maximum elongation. These scaling laws can be used for new reactor designs and for improving the performance of existing tokamak experiments. Our main results are as follows:\\
 \noindent (1) The maximum elongation is optimized when the triangularity of the wall is well matched by the effective plasma triangularity averaged over the total plasma volume to stabilize the $n=0$ mode effectively. The effective plasma triangularity increases with the Shafranov shift. Accordingly, as $\epsilon$, $\beta_p$ or $l_i$ increases, the Shafranov shift increases and $\delta_{opt}$ increases, as reflected in the scaling law for the optimal triangularity at the wall and plasma boundaries in Eq. (\ref{delta_res1}).\\ 
 (2) The sensitivity of $\kappa$ on $\delta$ is reduced by increasing $l_i$, as shown by the decrease of $\kappa_\delta$ in Eq. (\ref{C_res1}) and Figure 4. \\
 (3) Eq. (\ref{kappa0_res0}) and Eq. (\ref{kappa0_res1}) show that the dependence of $\kappa_0$ and $\kappa_1$ on the other physical parameters varies depending on the magnitude of the triangularity. Larger values for the triangularity typically result in smaller $\kappa_0$ and larger $\kappa_1$, i.e. a larger $\epsilon$ dependence.

 \newpage
 \clearpage
 
\appendix

 \section{Estimation of $\gamma \tau_w$ with moderate $l_i$ and $\beta_p$}\label{Appxb}

In Table A1, we recalculate the feedback parameter $\gamma \tau_w$ in [18] with the realistic values of the internal inductance $l_i\sim0.8$ and the poloidal beta $\beta_p \sim 1.0$ for several tokamaks.

\Table{\label{tab:gt}  The feedback paramter $\gamma \tau_w$ can be theoretically estimated by examining the data for several major large tokamak experiments from the references: ASDEX Upgrade (AUG) [24], Alcator C-Mod (C-Mod) [25], DIII-D [26], JET [27], NSTX [28], and ITER [29].  When information was missing regarding the plasma profiles, we assumed that the internal inductance $l_i$ and the poloidal beta $\beta_p$ were as given below.}

\br
 \centre{1}{Quantity}& \centre{6}{Device} \\
\br

	& AUG	& C-Mod&	DIII-D&	JET&	NSTX&	ITER\\
\br
Shot	&12145&	960214039&	73334&	49080&	132913&	---\\
 \mr
$\bar{p}[atm]$ &0.38&	1.02&	0.53&	0.42&	0.23&	1.73\\

 $a/R_0$ &0.51/1.60 &0.23/0.67 &0.61/1.67&0.91/2.91&0.58/0.86&2.00/6.20\\
$\epsilon$ &0.32	&0.34	&0.37	&0.31	&0.67	&0.32\\
 
$\kappa$ &1.84	 &1.77	&2.05	&1.93	&2.42	&1.72\\
 
$\delta$ &0.28	&0.70&	0.80&	0.36&	0.66&	0.33\\

$\Delta_i/a$ &0.17	&0.08&	0.11&	0.17&	0.16&	0.08\\
 
$b/a$ &1.17	&1.08&	1.11&	1.17&	1.16&	1.08\\
 
$\kappa_w$ &2.01	&1.86	&2.14	&2.09	&2.50	&1.81\\
 
$\kappa_w/\kappa$ &1.09	&1.05	&1.05	&1.08	&1.03	&1.06\\
 
$l_i$ &0.75	&0.75	&0.71	&0.72	&0.76	&0.78\\

$\beta_p$ &1.16&	0.90	&0.85	&0.86&	1.13&	1.20 \\

 \mr
$\gamma\tau_w$ &2.56 &1.06 &	1.59&	2.99 &	3.16 &	1.20 \\
      \br
\end{tabular}
\end{indented}
\end{table}

\end{document}